# Instagram photos reveal predictive markers of depression

Andrew G. Reece[a], Christopher M. Danforth[b,c]

**Abstract:** Using Instagram data from 166 individuals, we applied machine learning tools to successfully identify markers of depression. Statistical features were computationally extracted from 43,950 participant Instagram photos, using color analysis, metadata components, and algorithmic face detection. Resulting models outperformed general practitioners' average diagnostic success rate for depression. These results held even when the analysis was restricted to posts made before depressed individuals were first diagnosed. Photos posted by depressed individuals were more likely to be bluer, grayer, and darker. Human ratings of photo attributes (happy, sad, etc.) were weaker predictors of depression, and were uncorrelated with computationally-generated features. These findings suggest new avenues for early screening and detection of mental illness.

**Significance statement:** Photographs posted to Instagram can be analyzed using computational methods to screen for depression. Using only photographic details, such as color and brightness, our statistical model was able to predict which study participants suffered from depression, and performed better than the rate at which unassisted general practitioners typically perform during in-person patient assessments. Markers of depression were evident in Instagram posts even when only including posts made before the date of first depression diagnosis. These findings demonstrate how visual social media may be harnessed to make accurate inferences about mental health.

**Corresponding authors**: Andrew G. Reece, 33 Kirkland St, Cambridge, MA 02138, 617.395.6841, reece@g.harvard.edu; Christopher M. Danforth, 210 Colchester Ave, Burlington, VT, 05405, chris.danforth@uvm.edu

[a]Department of Psychology, Harvard University, Cambridge, MA 02138; [b]Computational Story Lab, Vermont Advanced Computing Core, and the Department of Mathematics and Statistics, University of Vermont, Burlington, VT 05401; [c]Vermont Complex Systems Center, University of Vermont, Burlington, VT 05401

The advent of social media presents a promising new opportunity for early detection and intervention in psychiatric disorders. Predictive screening methods have successfully analyzed online media to detect a number of harmful health conditions (1-11). All of these studies relied on text analysis, however, and none have yet harnessed the wealth of psychological data encoded in *visual* social media, such as photographs posted to Instagram. In this report, we introduce a methodology for analyzing photographic data from Instagram to predictively screen for depression.

There is good reason to prioritize research into Instagram analysis for health screening. Instagram members currently contribute almost 100 million new posts per day (12), and Instagram's rate of new users joining has recently outpaced Twitter, YouTube, LinkedIn, and even Facebook (13). A nascent literature on depression and Instagram use has so far either yielded results that are too general or too labor-intensive to be of practical significance for predictive analytics (14,15). In our research, we incorporated an ensemble of computational methods from machine learning, image processing, and other data-scientific disciplines to extract useful psychological indicators from photographic data. Our goal was to successfully identify and predict markers of depression in Instagram users' posted photographs.

> **Hypothesis 1:** Instagram posts made by individuals diagnosed with depression can be reliably distinguished from posts made by healthy individuals, using only measures extracted computationally from posted photos and associated metadata.

**Photographic markers of depression**

Photographs posted to Instagram offer a vast array of features that might be analyzed for psychological insight. The content of photographs can be coded for any number of characteristics: Are there people present? Is the setting in nature or indoors? Is it night or day? Image statistical properties can also be evaluated at a per-pixel level, including values for average color and brightness. Instagram metadata offers additional information: Did the photo receive any comments? How many "Likes" did it get? Finally, platform activity measures, such as usage and posting frequency, may also yield clues as to an Instagram user's mental state. We incorporated only a narrow subset of possible features into our predictive models, motivated in part by prior research into the relationship between mood and visual preferences.

In studies associating mood, color, and mental health, healthy individuals identified darker, grayer colors with negative mood, and generally preferred brighter, more vivid colors (16-19). By contrast, depressed individuals were found to prefer darker, grayer colors (17). In addition, Barrick, Taylor, & Correa (19) found a positive correlation between self-identification with depression and a tendency to perceive one's surroundings as gray or lacking in color. These findings motivated us to include measures of hue, saturation, and brightness in our analysis. We also tracked the use of Instagram filters, which allow users to modify the color and tint of a photograph.

Depression is strongly associated with reduced social activity (20,21). As Instagram is used to share personal experiences, it is reasonable to infer that posted photos with people in them may capture aspects of a user's social life. On this premise, we used a face detection algorithm to analyze Instagram posts for the presence and number of human faces in each photograph. We also counted the number of comments and likes each post received as measures of community engagement, and used posting frequency as a metric for user engagement.

**Early screening applications**

Hypothesis 1 is a necessary first step, as it addresses an unanswered basic question: Is depression detectable in Instagram posts? On finding support for Hypothesis 1, a natural question arises: Is depression detectable in Instagram posts, *before the date of first diagnosis*? After receiving a depression diagnosis, individuals may come to identify with their diagnosis (22, 23). Individuals' self-portrayal on social media may then be influenced by this identification. It is possible that a successful predictive model, trained on the entirety of depressed Instagram users' posting histories, might not actually detect depressive signals, per se, but rather purposeful content choices intended to convey a depressive condition. Training a model using only posts made prior to the date of first diagnosis addresses this potential confounding factor.

> **Hypothesis 2:** Instagram posts made by depressed individuals prior to the date of first clinical diagnosis can be reliably distinguished from posts made by healthy individuals.

If support is found for Hypothesis 2, this would not only demonstrate a methodological advance for researchers, but also serve as a proof-of-concept for future healthcare applications. As such, we benchmarked the accuracy of our model against the ability of general practitioners to correctly diagnose depression as shown in a meta-analysis by Mitchell, Vaze, and Rao (24). The authors analyzed 118 studies that evaluated general practitioners' abilities to correctly diagnose depression in their patients, without assistance from scales, questionnaires, or other measurement instruments. Out of 50,371 patient outcomes included across the pooled studies, 21.9% were actually depressed, as evaluated separately by psychiatrists or validated interview-based measures conducted by researchers. General practitioners were able to correctly rule out depression in non-depressed patients 81% of the time, but only diagnosed depressed patients correctly 42% of the time. We refer to these meta-analysis findings (24) as a comparison point to evaluate the usefulness of our models.

A major strength of our proposed models is that their features are generated using entirely computational means - pixel analysis, face detection, and metadata parsing - which can be done at scale, without additional human input. It seems natural to wonder whether these machine-extracted features pick up on similar signals that humans might use to identify mood and psychological condition, or whether they attend to wholly different information. A computer may be able to analyze the average saturation value of a million pixels, but can it pick out a happy selfie from a sad one? Are machine learning and human opinion sensitive to the same indicators of depression? Furthermore, insight into these issues may help to frame our results in the larger discussion around human versus machine learning, which occupies a central role in the contemporary academic landscape.

To address these questions, we solicited human assessments of the Instagram photographs we collected. We asked new participants to evaluate photos on four simple metrics: happiness, sadness, interestingness, and likability. These ratings categories were intended to capture human impressions that were both intuitive and quantifiable, and which had some relationship to established depression indicators. DSM-IV (20) criteria for Major Depressive Disorder includes feeling sad as a primary criterion, so sadness and happiness seemed obvious candidates as ratings categories. Epstein et al. (25) found depressed individuals "had difficulty reconciling a self-image as an 'outgoing likeable person'", which prompted likability as an informative metric. We hypothesized that human raters should find photographs posted by depressed individuals to be sadder, less happy, and less likable, on average. Finally, we considered interestingness as a novel factor, without a clear directional hypothesis.

> **Hypothesis 3a:** Human ratings of Instagram posts on common semantic categories can distinguish between posts made by depressed and healthy individuals.
> **Hypothesis 3b**: Human ratings are positively correlated with computationally-extracted features.

If human and machine[1] predictors show positive correlation, we can infer that each set of features reveals similar indicators of depression. In this case, the strength of the human model simply suggests whether it is better or worse than the machine model. On the other hand, if machine and human features show little or no correlation, then regardless of human model performance, we would know that the machine features are capable of screening for depression, but use different information signals than what are captured by the affective ratings categories.

---

[1] The term "machine" (eg. "machine predictors", "machine model") is used as shorthand for the computational feature extraction we employed. Significant human biases informed this process, as the initial selection of features for extraction involved entirely human decision-making.

# Method

**Data Collection**

Data collection was crowdsourced using Amazon's Mechanical Turk (MTurk) crowdwork platform[2]. Separate surveys were created for depressed and healthy individuals. In the depressed survey, participants were invited to complete a questionnaire that involved passing a series of inclusion criteria, responding to a standardized clinical depression survey, answering questions related to demographics and history of depression, and sharing social media history. We used the CES-D (Center for Epidemiologic Studies Depression Scale) questionnaire to screen participant depression levels (26). CES-D assessment quality has been demonstrated to be on-par with other depression inventories, including the Beck Depression Inventory and the Kellner Symptom Questionnaire (27, 28). Healthy participants were screened to ensure they were active users of Instagram, and had no history of depression.

Qualified participants were asked to share their Instagram usernames and history. An app embedded in the survey allowed participants to securely log into their Instagram accounts and agree to share their data. Upon securing consent, we made a one-time collection of participants' entire Instagram posting history. In total we collected 43,950 photographs from 166 Instagram users.

We asked a different set of MTurk crowdworkers to rate the Instagram photographs collected. This new task asked participants to rate a random selection of 20 photos from the data we collected. Raters were asked to judge how interesting, likable, happy, and sad each photo seemed, on a continuous 0-5 scale. Each photo was rated by at least three different raters, and ratings were averaged across raters[3]. Raters were not informed that photos were from Instagram, nor were they given any information about the study participants who provided the photos, including mental health status. Each ratings category showed good inter-rater agreement.

Only a subset of participant Instagram photos were rated (N=13,184). We limited ratings data to a subset because this task was time-consuming for crowdworkers, and so proved a costly form of data collection. For the depressed sample, ratings were only made for photos posted within a year in either direction of the date of first depression diagnosis. Within this subset, for each user the nearest 100 posts prior to the diagnosis date were rated. For the healthy sample, the most recent 100 photos from each user's date of participation in this study were rated.

**Participant safety and privacy**

Data privacy was a concern for this study. Strict anonymity was nearly impossible to guarantee to participants, given that usernames and personal photographs posted to Instagram often contain identifiable features. We made sure participants were informed of the risks of being personally identified, and assured them that no data with personal identifiers, including usernames, would be made public or published in any format.

---

[2] See SI Appendix I for details on data collection procedures, and SI Appendix III for summary statistics.
[3] See SI Appendix X for measures of inter-rater agreement.

**Improving data quality**

We employed several quality assurance measures in our data collection process to reduce noisy and unreliable data. Our surveys were only visible to MTurk crowdworkers who had completed at least 100 previous tasks with a minimum 95% approval rating; MTurk workers with this level of experience and approval rating have been found to provide reliable, valid survey responses (29). We also restricted access to only American IP addresses, as MTurk data collected from outside the United States are generally of poorer quality (30). All participants were only permitted to take the survey once.

We excluded participants who had successfully completed our survey, but who had a lifetime total of fewer than five Instagram posts. We also excluded participants with CES-D scores of 21 or lower. Studies have indicated that a CES-D score of 22 represents an optimal cutoff for identifying clinically relevant depression across a range of age groups and circumstances (31,32).

**Feature extraction**

Several different types of information were extracted from the collected Instagram data. We used total posts per user, per day, as a measure of user activity. We gauged community reaction by counting the number of comments and "likes" each posted photograph received. Face detection software was used to determine whether or not a photograph contained a human face, as well as count the total number of faces in each photo, as a proxy measure for participants' social activity levels[4]. Pixel-level averages were computed for Hue, Saturation, and Value (HSV), three color properties commonly used in image analysis. Hue describes an image's coloring on the light spectrum (ranging from red to blue/purple). Lower hue values indicate more red, and higher hue values indicate more blue. Saturation refers to the vividness of an image. Low saturation makes an image appear grey and faded. Value refers to image brightness. Lower brightness scores indicate a darker image. See Fig. 1 for a comparison of high and low HSV values. We also checked metadata to assess whether an Instagram-provided filter was applied to alter the appearance of a photograph. Collectively, these measures served as the feature set in our primary model. For the separate model fit on ratings data, we used only the four ratings categories (happy, sad, likable, interesting) as predictors.

---

[4] See Appendix II for face detection algorithm details.

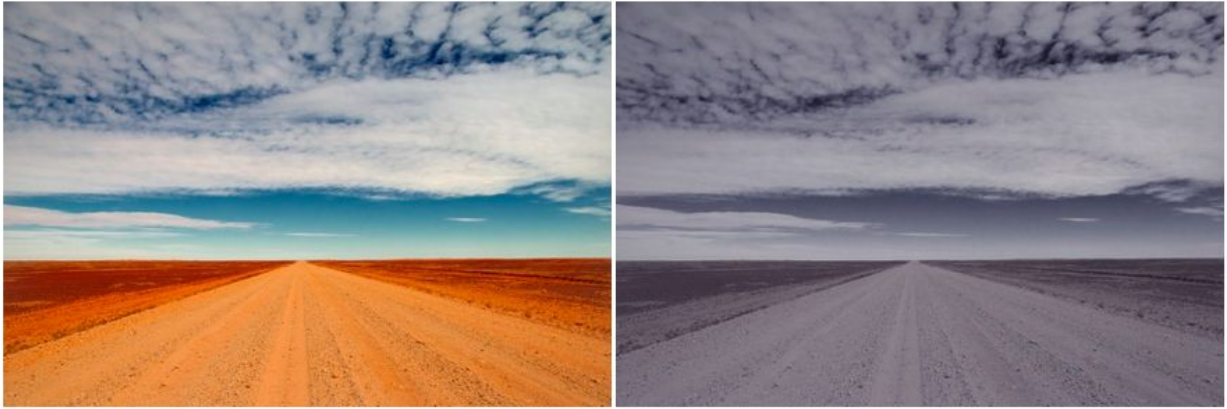

Fig. 1. Comparison of HSV values. Right photograph has higher Hue (bluer), lower Saturation (grayer), and lower Brightness (darker) than left photograph. Instagram photos posted by depressed individuals had HSV values shifted towards those in the right photograph, compared with photos posted by healthy individuals.

**Units of observation**

In determining the best time span for this analysis, we encountered a difficult question: When and for how long does depression occur? A diagnosis of depression does not indicate the persistence of a depressive state for every moment of every day, and to conduct analysis using an individual's entire posting history as a single unit of observation is therefore rather specious. At the other extreme, to take each individual photograph as units of observation runs the risk of being too granular. DeChoudhury et al. (5) looked at all of a given user's posts in a single day, and aggregated those data into per-person, per-day units of observation. We adopted this precedent of "user-days" as a unit of analysis[5].

**Statistical framework**

We used Bayesian logistic regression with uninformative priors to determine the strength of individual predictors. Two separate models were trained. The All-data model used all collected data to address Hypothesis 1. The Pre-diagnosis model used all data collected from healthy participants, but only pre-diagnosis data from depressed participants, to address Hypothesis 2. We also fit an "intercept-only" model, in which all predictors are zero-weighted to simulate a model under a null hypothesis. Bayes factors were used to assess model fit. Details on Bayesian estimation, model optimization, and diagnostic checks are available in SI Appendix IV, V, and VII. Frequentist regression output is provided for comparison in Appendix VI.

We also employed a suite of supervised machine learning algorithms to estimate the predictive capacity of our models. We report prediction results only from the best-performing algorithm, a 1200-tree Random Forests classifier. As an informal benchmark for comparison, we present general practitioners' unassisted diagnostic accuracy as reported in Mitchell, Vaze, and Rao (MVR) (24)[6].

---

[5] Occasionally, when reporting results we refer to "observations" as "participants", eg. "depressed participants received fewer likes". It would be more correct to use the phrase "photographic data aggregated by participant-user-days" instead of "participants". We chose to sacrifice a degree of technical correctness for the sake of clarity.

# Results

Both All-data and Pre-diagnosis models were decisively superior to a null model ($K_{All}$ = 157.5; $K_{Pre}$ = 149.8)[7]. All-data predictors were significant with 99% probability. Pre-diagnosis and All-data confidence levels were largely identical, with two exceptions: Pre-diagnosis Brightness decreased to 90% confidence, and Pre-diagnosis posting frequency dropped to 30% confidence, suggesting a null predictive value in the latter case.

Increased hue, along with decreased brightness and saturation, predicted depression. This means that photos posted by depressed individuals tended to be bluer, darker, and grayer (see Fig. 2). The more comments Instagram posts received, the more likely they were posted by depressed participants, but the opposite was true for likes received. In the All-data model, higher posting frequency was also associated with depression. Depressed participants were more likely to post photos with faces, but had a lower average face count per photograph than healthy participants. Finally, depressed participants were less likely to apply Instagram filters to their posted photos.

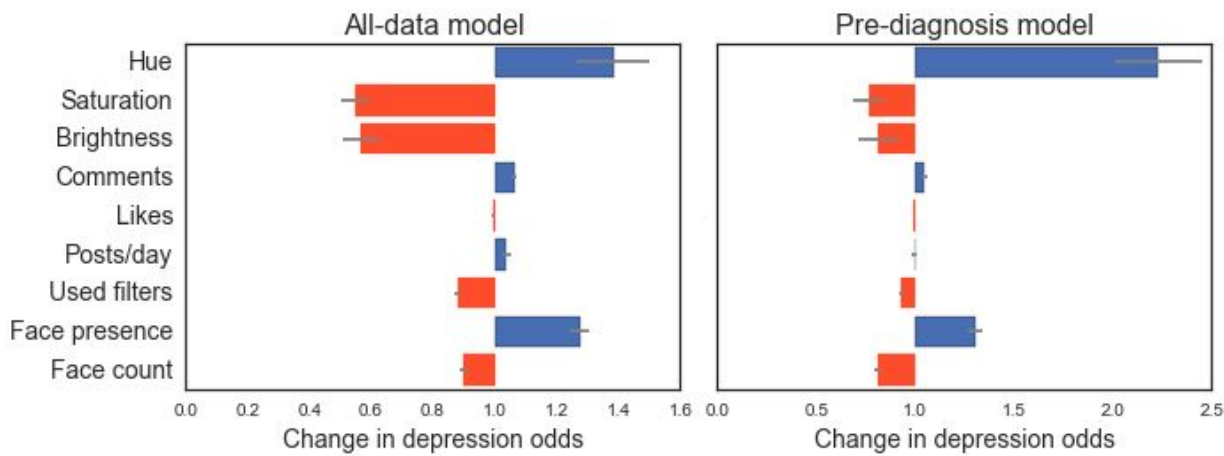

Fig. 2. Magnitude and direction of regression coefficients in All-data (N=24,713) and Pre-diagnosis (N=18,513) models. X-axis values represent the adjustment in odds of an observation belonging to depressed individuals, per standardized unit increase of each predictive variable. Odds were generated by exponentiating logistic regression log-odds coefficients.

---

[6] Comparing point estimates of accuracy metrics is not a statistically robust means of model comparison. However, we felt it was more meaningful to frame our findings in a realistic context, rather than to benchmark against a naive model that simply predicted the majority class for all observations.
[7] $K$ abbreviates the Bayes Factor ratio between the subscripted model and a null model. See SI Appendix IV for $K$ value legend.

A closer look at filter usage in depressed versus healthy participants provided additional texture. Instagram filters were used differently by depressed and healthy individuals ($\chi2_{All} = 907.84$, $p = 9.17e-164$; $\chi2_{Pre} = 813.80$, $p = 2.87e-144$). In particular, depressed participants were less likely than healthy participants to use any filters at all. When depressed participants did employ filters, they most disproportionately favored the "Inkwell" filter, which converts color photographs to black-and-white images. Conversely, healthy participants most disproportionately favored the Valencia filter, which lightens the tint of photos. Examples of filtered photographs are provided in SI Appendix VIII.

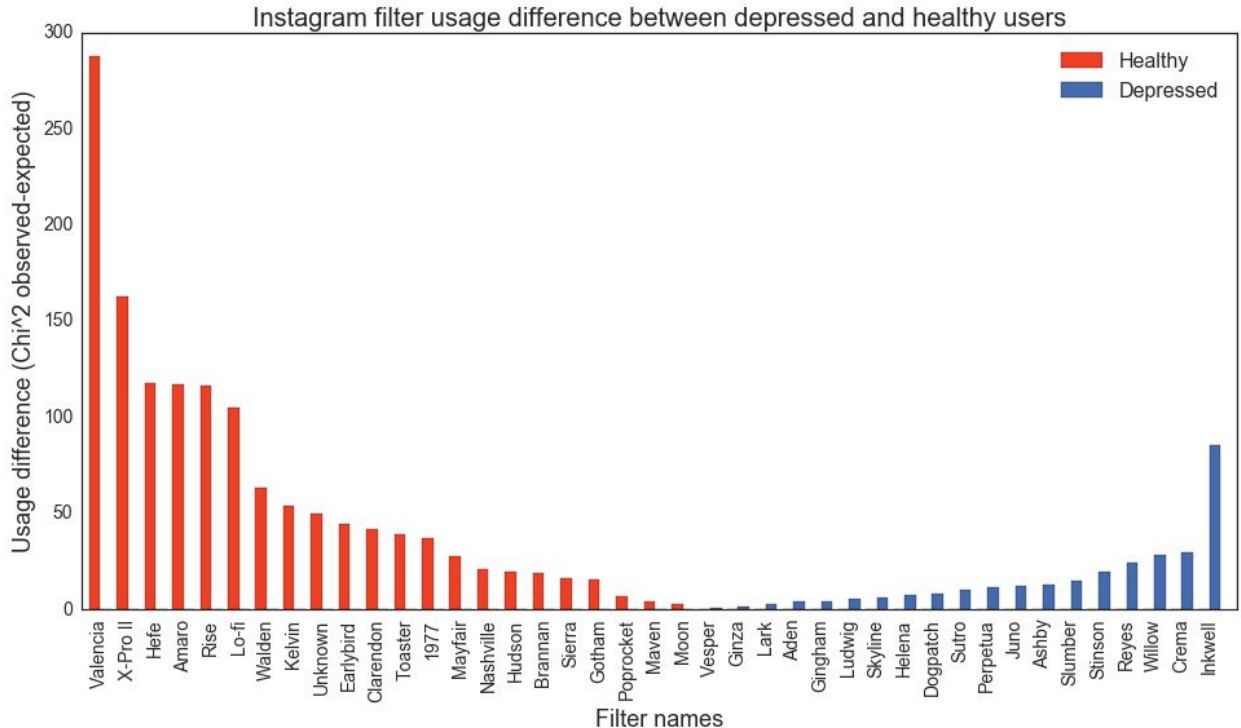

Fig. 3. Instagram filter usage among depressed and healthy participants. Bars indicate difference between observed and expected usage frequencies, based on a Chi-squared analysis of independence. Blue bars indicate disproportionate use of a filter by depressed compared to healthy participants, orange bars indicate the reverse. All-data model results are displayed, similar results were observed for the Pre-diagnosis model, see SI Appendix XI.

Our best All-data machine learning classifier, averaged over five randomized iterations, improved over MVR general practitioner accuracy on most metrics. Compared with MVR results, the All-data model was less conservative (lower specificity) but better able to positively identify observations from depressed individuals (higher recall). Given 100 observations, our model correctly identified 70% of all depressed cases (n=37), with a relatively low number of false alarms (n=23) and misses (n=17).

Pre-diagnosis predictions showed improvement over the MVR benchmark on precision and specificity. The Pre-diagnosis model found only about a third of actual depressed observations, but it was correct most of the time when it did assign a depressed label. By comparison, although MVR general practitioners discovered more true cases of depression, they were more likely than not to misdiagnose healthy subjects as depressed.

|             | MVR μ | All-data μ (σ) | Pre-diagnosis μ (σ) |
|---|---|---|---|
| Recall      | .510  | .697 (.008)    | .318 (.012)         |
| Specificity | .813  | .478 (.012)    | .833 (.010)         |
| Precision   | .42   | .604 (.009)    | .541 (.009)         |
| NPV         | .858  | .579 (.008)    | .665 (.006)         |
| F1          | .461  | .647 (.003)    | .401 (.008)         |

Table 1. Comparison of accuracy metrics for All-data and Pre-diagnosis model predictions. General practitioners' diagnostic accuracy from Mitchell, Vaze, & Rao (25) (MVR) is included for comparison.

Out of the four predictors used in the human ratings model (happiness, sadness, likability, interestingness), only the sadness and happiness ratings were significant predictors of depression. Depressed participants' photos were more likely to be sadder and less happy than those of healthy participants. Ratings assessments generally showed strong patterns of correlation with one another, but exhibited extremely low correlation with computational features. The modest positive correlation of human-rated happiness with the presence and number of faces in a photograph was the only exception to this trend. Correlation matrices for all models are available in Appendix IX.

## Discussion

The present study employed machine learning techniques to screen for depression using photographs posted to Instagram. Our results supported Hypothesis 1, that markers of depression are observable in Instagram user behavior, and Hypothesis 2, that these depressive signals are detectable in posts made even before the date of first diagnosis. Human ratings proved capable of distinguishing between Instagram posts made by depressed and healthy individuals (Hypothesis 3a), but showed little or no correlation with most computational features (Hypothesis 3b). Our findings establish that visual social media data are amenable to analysis of affect using scalable, computational methods. One avenue for future research might integrate textual analysis of Instagram posts' comments, captions, and tags. Considering the early success of textual analysis in detecting psychological signals on social media (5,33), the modeling of textual and visual features together could prove superior to either medium on its own.

Our model showed considerable improvement over the ability of unassisted general practitioners to correctly diagnose depression. On average, more than half of general practitioners' depression diagnoses were false positives (24). By comparison, the majority of both All-data and Pre-diagnosis depression classifications were correct. As false diagnoses are costly for both healthcare programs and individuals, this improvement is noteworthy. Health care providers may be able to improve quality of care and better identify individuals in need of treatment based on the simple, low-cost methods outlined in this report. Given that mental health services are unavailable or underfunded in many countries (34), this computational approach, requiring only patients' digital consent to share their social media histories, may open avenues to care which are currently difficult or impossible to provide.

      Although our Pre-diagnosis prediction engine was rather conservative, and tended to classify most observations as healthy, its accuracy likely represents a lower bound on performance. Ideally, we would have used the All-data classifier to evaluate the Pre-diagnosis data, as the All-data model was trained on a much larger dataset. Since the Pre-diagnosis data constituted a subset of the full dataset, applying the All-data model to Pre-diagnosis observations would have artificially inflated accuracy, due to information leakage between training and test data. Instead, we trained a separate classifier using training and test partitions contained within the Pre-diagnosis data. This left the Pre-diagnosis model with considerably fewer data points to train on, resulting in weaker predictive power. As a result, it is likely that Pre-diagnosis accuracy scores understate the method's true capacity.

      Regarding the strength of specific predictive features, some results matched common perceptions regarding the effects of depression on behavior. Photos posted to Instagram by depressed individuals were more likely to be bluer, grayer, darker, and receive fewer likes. Depressed Instagram users in our sample had an outsized preference for filtering out all color from posted photos, and showed an aversion to artificially lightening photos, compared to healthy users. These results are congruent with the literature linking depression and a preference for darker, bluer, and monochromatic colors (16-19).

      Other, seemingly intuitive, relationships failed to emerge. For example, the sadness of a photo, and the extent to which it is bluer, darker, and grayer than other photos, seem like close semantic matches. Despite both being strong predictors of depression, however, sadness ratings and color were statistically unrelated.

      Algorithmic face detection yielded intriguing results. Depressed users were more likely to post photos with faces, but they tended to post fewer faces per photo. Fewer faces may be an oblique indicator that depressed users interact in smaller social settings, which would be in accordance with research linking depression to reduced social interactivity (5,20,21). That depressed Instagram users posted more photos with faces overall, however, offers less clear interpretation. Depressed individuals have been shown to use more self-focused language (35), and it may be that this self-focus extends to photographs, as well. If so, it may be that the abundance of low-face-count photos posted by depressed users are, in fact, self-portraits. This "sad selfie" hypothesis remains untested.

      A limitation of these findings concerns the non-specific use of the term "depression" in the data collection process. We acknowledge that depression describes a general clinical state,

and is frequently comorbid with other conditions. It is possible that a specific diagnostic class is responsible for driving the observed results, and subsequent research should adjust questionnaires to acquire specific diagnostic information. Additionally, it is possible that our results are specific to individuals who received clinical diagnoses. Current perspectives on depression treatment indicate that people who are "well-informed and psychologically minded, experience typical symptoms of depression and little stigma, and have confidence in the effectiveness of treatment, few concerns about side effects, adequate social support, and high self-efficacy" seek out mental health services (25). The intersection of these qualities with typical Instagram user demographics suggests caution in making broad inferences about depression, based on our findings.

As the methods employed in this research provide a tool for inferring personal information about individuals, two points of caution should be considered. First, data privacy and ethical research practices are of particular concern, given recent admissions that individuals' social media data were experimentally manipulated or exposed without permission (36,37). It is perhaps reflective of a current general skepticism towards social media research that, of the 509 individuals who began our survey, 221 (43%) refused to share their Instagram data, even after we provided numerous privacy guarantees. Future research should prioritize establishing confidence among experimental participants that their data will remain secure and private. Second, data trends often change over time, leading socio-technical models of this sort to degrade without frequent calibration (38). The findings reported here should not be taken as enduring facts, but rather as a methodological foundation upon which to build and refine subsequent models.

Paired with a commensurate focus on upholding data privacy and ethical analytics, the present work may serve as a blueprint for effective mental health screening in an increasingly digitalized society. More generally, these findings support the notion that major changes in individual psychology are transmitted in social media use, and can be identified via computational methods.

**Acknowledgments**. We thank K. Lix for conversations and manuscript review. A.R. was supported by the Sackler Scholar Programme in Psychobiology. C.D. was supported by NSF grant #1447634.

# Supplementary Information

## I. Data collection procedures

This study was reviewed and approved by the Harvard University Institutional Review Board, approval #15-2529 and by the University of Vermont Institutional Review Board, approval #CHRMS-16-135. All study participants were informed of and acknowledged all study goals, expectations, and procedures, including data privacy procedures, prior to any data collection. Surveys were built using the Qualtrics survey platform. Analyses were conducted using the Python and R programming languages. Social media data collection apps were written in Python, using the Instagram developer's Application Programming Interface (API).

The survey for depressed participants collected age data from participants, and asked qualified participants questions related to their first depression diagnosis and social media usage at that time. These questions were given in addition to the CES-D scale. The purpose of these questions was to determine:
- The date of first depression diagnosis
- Whether or not the individual suspected being depressed before diagnosis, and,
- If so, the number of days prior to diagnosis that this suspicion began

In the case that participants could not recall exact dates, they were instructed to approximate the actual date.

The survey for healthy participants collected age and gender data from participants. It also asked four questions regarding personal health history, which were used as inclusion criteria for this and three other studies. These questions were as follows:
- Have you ever been pregnant?
- Have you ever been clinically diagnosed with depression?
- Have you ever been clinically diagnosed with Post-Traumatic Stress Disorder?
- Have you ever been diagnosed with cancer?

Participants' responses to these questions were not used at all in analysis, and only served to include qualified respondents in each of the various studies, including the depression-related study reported here.

II. Face Detection

We used an elementary face detection script, based on an open source demonstration (https://gist.github.com/dannguyen/cfa2fb49b28c82a1068f). The main adjustment we made from the open source demo was to run through the detection loop twice, using two differing scale factors. A single scale factor had difficulty finding both small and large faces.
Parameters used: scale_factors = [1.05, 1.4], min_neighbors = 4, min_size = (20px,20px)
Algorithm accuracy was assessed by manually coding a random sample of 400 photos (100 photos from each of combination of depressed/healthy, detected/undetected). Detection accuracy was roughly equal across groups:

Face detection accuracy:
    Depressed, No face detected: 77% accurate
    Healthy, No face detected: 79% accurate

    Depressed, 1+ faces detected: 59% accurate
    Healthy, 1+ faces detected: 61% accurate

The mean difference in counted faces (detected faces minus actual faces), indicated that the algorithm slightly undercounted the number of faces in photos, for both depressed participants ($\mu = -0.015$, $\sigma = 1.21$) as well as healthy participants ($\mu = -0.215$, $\sigma = 2.07$). In both groups, the algorithm undercounted by less than a single face, on average.

III. Summary statistics

All data collection took place between February 1, 2016 and April 6, 2016. Across both depressed and healthy groups, we collected data from 166 Instagram users, and analyzed 43,950 posted photographs. The mean number of posts per user was 264.76 (SD=396.06). This distribution was skewed by a smaller number of frequent posters, as evidenced by a median value of just 122.5 posts per user.

Among depressed participants, 84 individuals successfully completed participation and provided access to their Instagram data. Imposing the CES-D cutoff reduced the number of viable participants to 71. The mean age for viable participants was 28.8 years (SD=7.09), with a range of 19 to 55 years. Dates of participants' first depression diagnoses ranged from February 2010 to January 2016, with nearly all diagnosis dates (90.1%) occurring in the period 2013-2015.

Among healthy participants, 95 participants completed participation and provided access to their Instagram data. The mean age for this group was 30.7 years, with a range of 19 to 53 years, and 65.3% of respondents were female. (Gender data were not collected for the depressed sample.)

All-data model data consisted of participants' entire Instagram posting histories, consisted of 43,950 Instagram posts (24,811 depressed) over 166 individuals (71 depressed). Aggregation by user-days compressed into 24,713 observations (13,230 depressed). Observations from depressed participants accounted for 53.4% of the entire dataset.

Pre-diagnosis model data used only Instagram posts from depressed participants made prior to the date of first depression diagnosis, along with the same full dataset from healthy participants as used in the All-data model. These data consisted of a total of 32,311 posts (13,192 depressed). There were 18,513 aggregated-unit observations in total (7,030 depressed). Observations from depressed participants accounted for 38% of this dataset.

|  | Users | Posts | Posts ($\mu$) | Posts ($\sigma$) | Posts (median) |
|---|---|---|---|---|---|
| Total | 166 | 43,950 | 264.76 | 396.06 | 122.5 |
| Depressed | 71 | 24,811 | 349.45 | 441.19 | 196.0 |
| Healthy | 95 | 19,139 | 201.46 | 347.76 | 100.0 |

Table S1. Summary statistics for data collection (N=43,950).

## IV. Statistical framework

*Bayesian logistic regression*

A Bayesian framework avoids many of the inferential challenges of frequentist null hypothesis significance testing, including reliance on p-values and confidence intervals, both of which are subject to frequent misuse and misunderstanding (39-42). For comparison, results from frequentist logistic regression output are included below; both methods are largely in agreement.

Logistic regression was conducted using the `MCMClogit` function from the R package `MCMCpack` (43). This function asserts a model of the following form :

$$y_i \sim Bernoulli(\pi_i)$$

With the inverse link function:

$$\pi_i = \frac{\exp(x_i'\beta)}{1 + \exp(x_i'\beta)}$$

And a multivariate Normal prior on $\beta$:

$$\beta \sim \mathcal{N}(b_0, B_0^{-1})$$

We selected "uninformative" priors for all parameters in $\beta$, with $b_0 = 0, B_0 = 0.0001$. While generally it is preferable to specify Bayesian priors, in this setting our parameters of interest were entirely novel, and so were not informed by prior literature or previous testing.

The `MCMClogit()` function employs a Metropolis algorithm to perform Markov Chain Monte Carlo (MCMC) simulations. The Instagram model simulation used two MCMC chains of 100,000 iterations with a burn-in of 10,000 and no thinning. The use of thinning for achieving higher-precision estimates from posterior samples is questionable when compared to simply running longer chains (44). While no best practice has been established for how long an unthinned chain should be, Christensen et al. (45) advised: "Unless there is severe autocorrelation, e.g., high correlation with, say [lag]=30, we don't believe that thinning is worthwhile". In our MCMC chains, we observed low autocorrelation at a lag of 30, and so felt confident in foregoing thinning. For comparison, we also ran a 100,000-iteration chain, thinned to every 10th iteration, with a burn-in of 5,000. While autocorrelation was noticeably reduced at shorter lags, this chain yielded near-identical parameter estimates from the posterior.

Recall that Bayesian regression coefficients are not assigned p-values or any other significance measures conventional in frequentist null-hypothesis significance testing (NHST). We have provided Highest Posterior Density Intervals (HPDIs) for the highest probability at which the interval excludes zero as a possible coefficient value. For example, if a 99% HPDI is reported, it means that, based on averaged samples from the simulated joint posterior distribution, the coefficient in question has a 99% probability of being non-zero. References to variable "significance" in the Results section relate only to the probability that a variable's parameter estimate is non-zero, eg. "Variable X was significant with 99% probability".

Bayes factors were used to assess model fit. Given two models $M_a$, $M_b$ parameterized by parameter vectors $\theta_a$, $\theta_b$, and data $D$, the Bayes factor is computed as the ratio

$$K = \frac{Pr(D|\theta_a)}{Pr(D|\theta_b)} = \frac{\int Pr(\theta_a|M_a)Pr(D|\theta_a, M_a)\,d\theta_a}{\int Pr(\theta_b|M_b)Pr(D|\theta_b, M_b)\,d\theta_b}$$

A positive-valued Bayes factor supports model $M_a$ over $M_b$. Jeffreys (46) established the following key for interpreting $K$ in terms of evidence for $M_a$ as the stronger model:

$K < 10^0$: Negative evidence (supports $M_b$)
$10^0 < K < 10^{1/2}$: Barely worth mentioning
$10^{1/2} < K < 10^1$: Substantial
$10^1 < K < 10^{3/2}$: Strong
$10^{3/2} < K < 10^2$: Very strong
$K > 10^2$: Decisive

Markov Chain Monte Carlo (MCMC) chains showed good convergence across all estimated parameters on every fitted model. In all models, Gelman-Rubin diagnostics (47) indicated simulation chain convergence, with point estimates of 1.0 for each parameter. Geweke diagnostics (48) also indicated post-burn-in convergence. Autocorrelation was observed within acceptable levels. Trace, density, and autocorrelation plots for all models are presented in SI Appendix V.

*Machine learning models*
We employed a suite of supervised machine learning algorithms to estimate the predictive capacity of our models. In a supervised learning paradigm, parameter weights are determined by training on a labeled subset of the total available data ("labeled" here means that the response classes are exposed). Fitted models are then used to predict class membership for each observation in the remaining unlabeled "holdout" data. All of our machine learning classifiers were trained on a randomly-selected 70% of total observations, and tested on the remaining 30%. We employed stratified five-fold cross-validation to optimize hyperparameters, and averaged final model output metrics over five separate randomized runs.

Random Forests parameters were optimized using stratified five-fold cross-validation. The optimization routine traversed every combination over the following values (best performing values are highlighted above in red):

    n_estimators = [120, 300, 500, 800, 1200]
    max_depth = [5, 8, 15, 25, 30, None]
    min_samples_split = [1, 2, 5, 10, 15, 100]
    min_samples_leaf = [1, 2, 5, 10]
    max_features = ['log2', 'sqrt', None]

In evaluating binary classification accuracy, a simple proportion of correct classifications ("naive accuracy") is often inappropriate. In cases where data exhibit a class imbalance, i.e. more healthy than depressed observations (or vice-versa), reporting naive accuracy can be misleading. (A classification accuracy of 95% seems excellent until it is revealed that 95% of the data modeled belong to a single class.) Additionally, naive accuracy scores are opaque to the specific strengths and weaknesses of a binary classifier. Instead, we report precision, recall, specificity, negative predictive value, and F1 scores for fuller context.

## V. MCMC Diagnostics

### All-data model

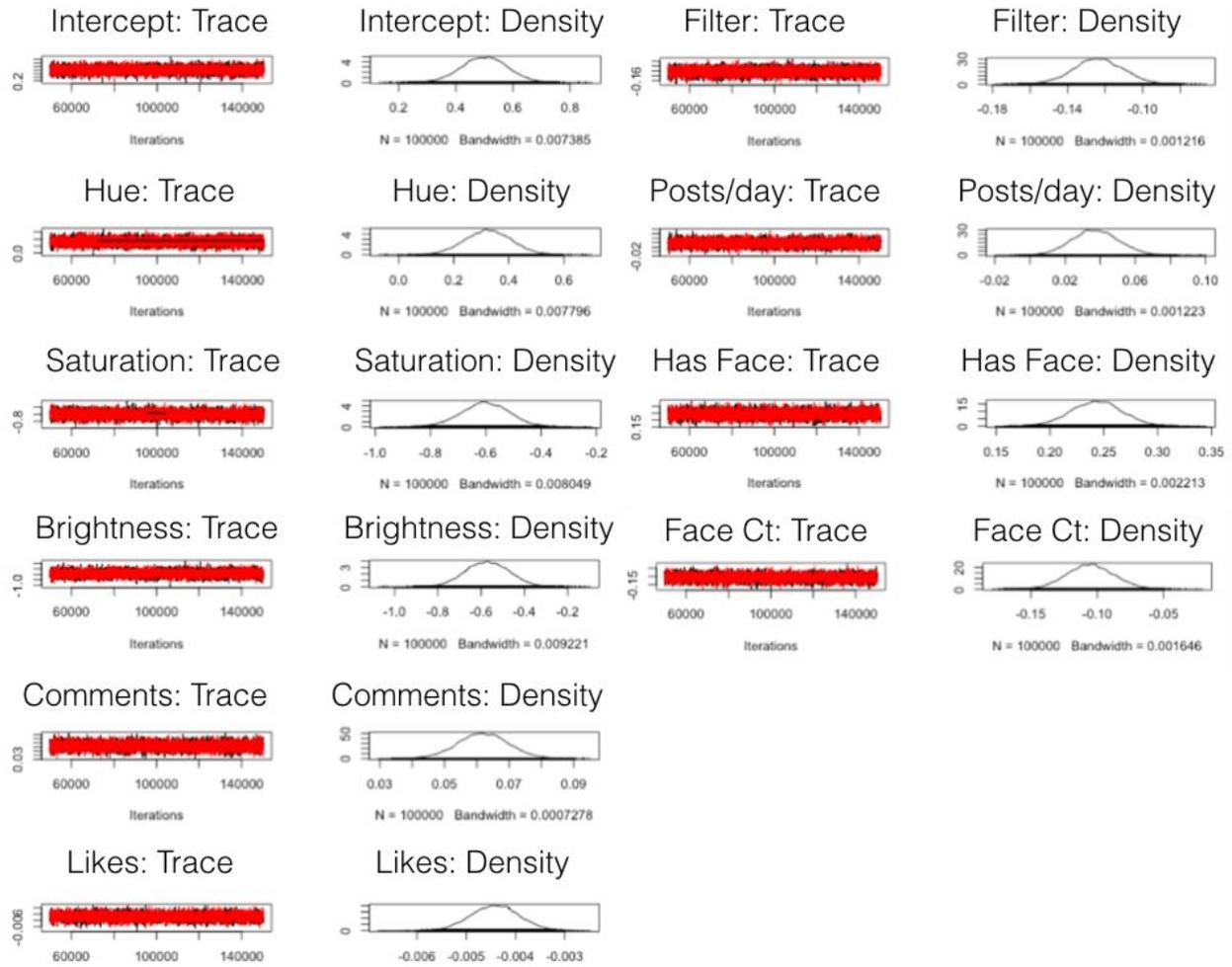

Fig. S1. Trace and density plots for All-data model MCMC simulations.

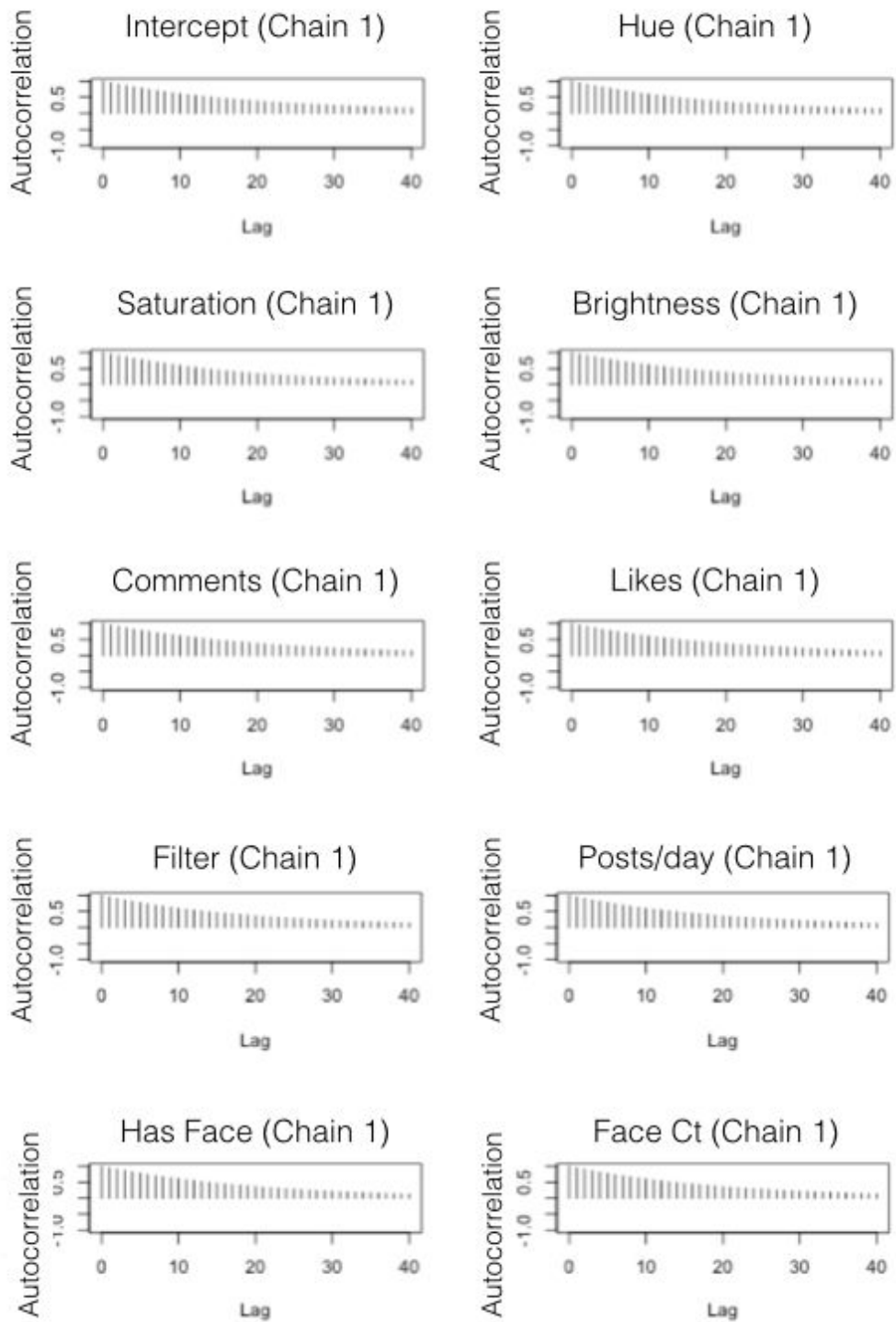

Fig. S2. Autocorrelation plot for All-data model MCMC simulations. First chain only is displayed for conciseness (second chain output is nearly identical).

Pre-diagnosis model

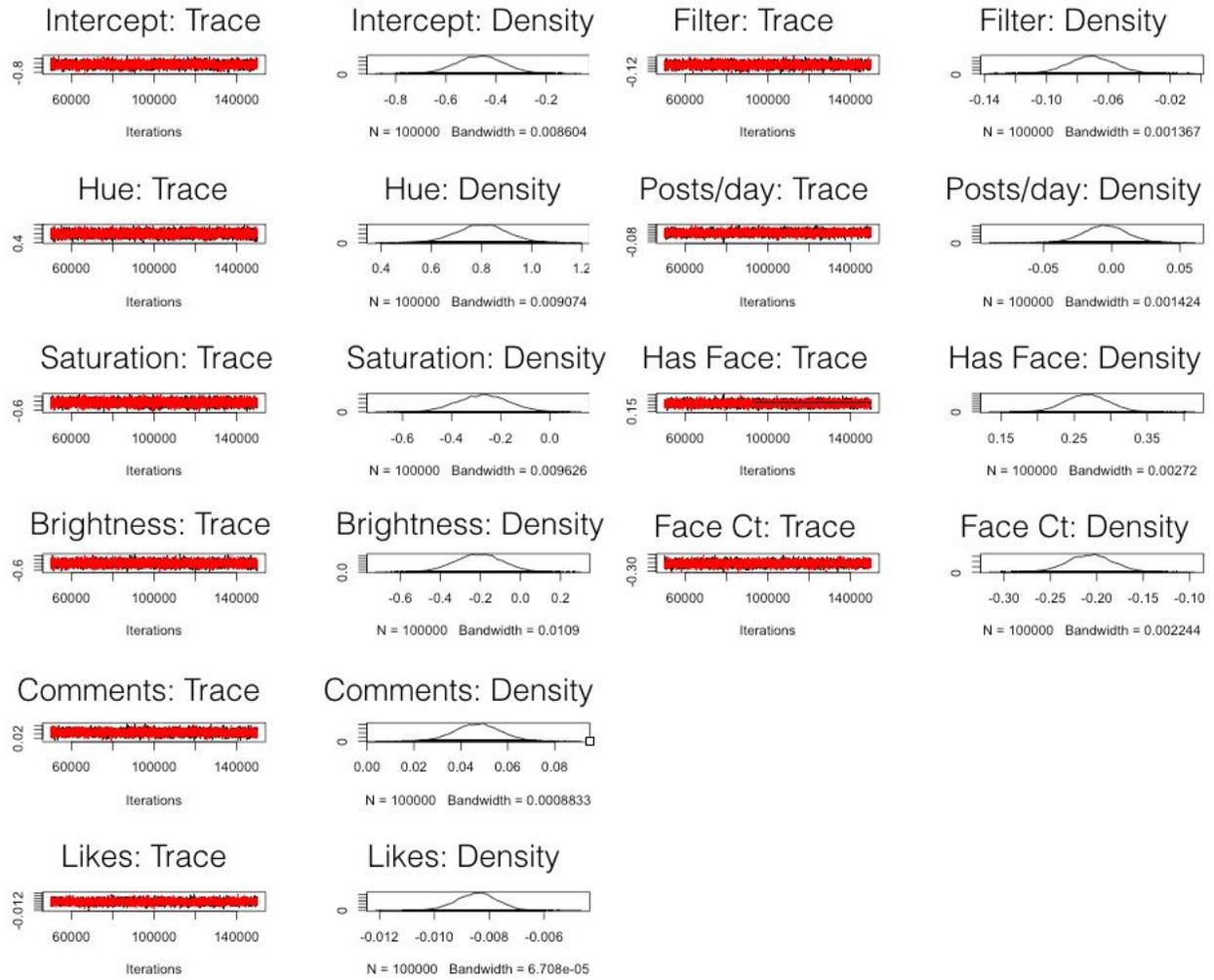

Fig. S3. Trace and density plots for Pre-diagnosis model MCMC simulations.

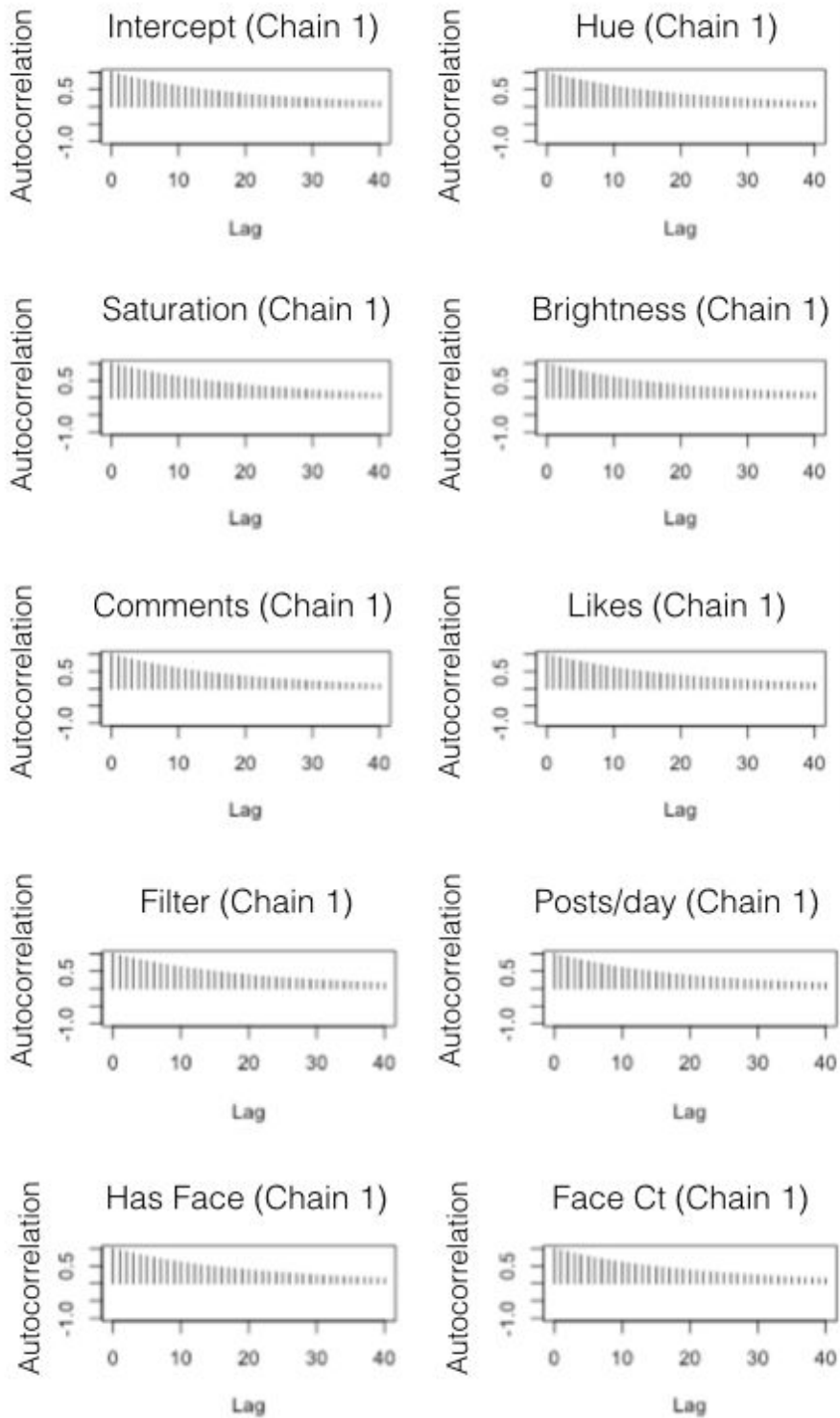

Fig. S4. Autocorrelation plot for Pre-diagnosis model MCMC simulations. First chain only is displayed for conciseness (second chain output is nearly identical).

Ratings model

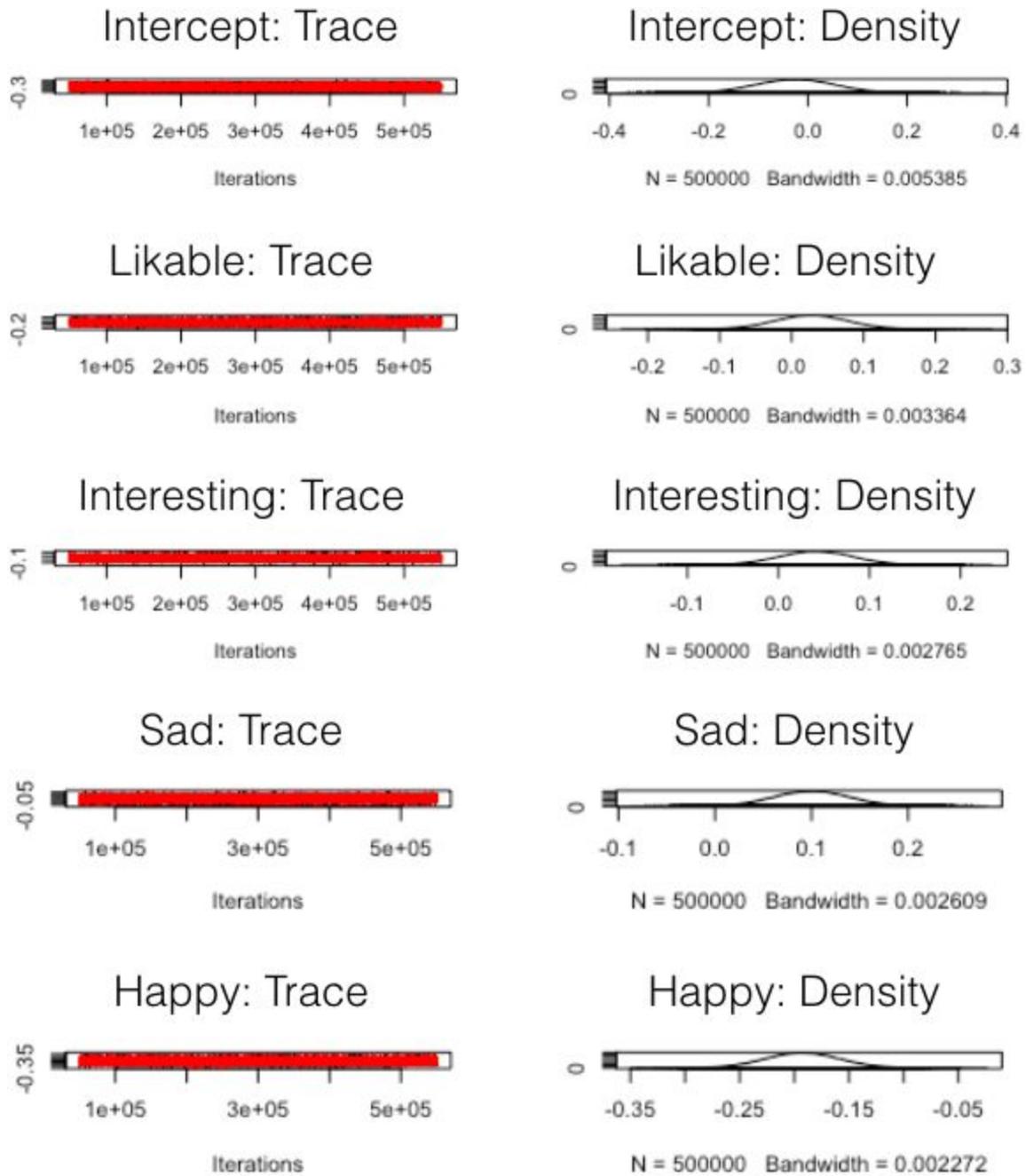

Fig. S5. Trace and density plots for Ratings model MCMC simulations.

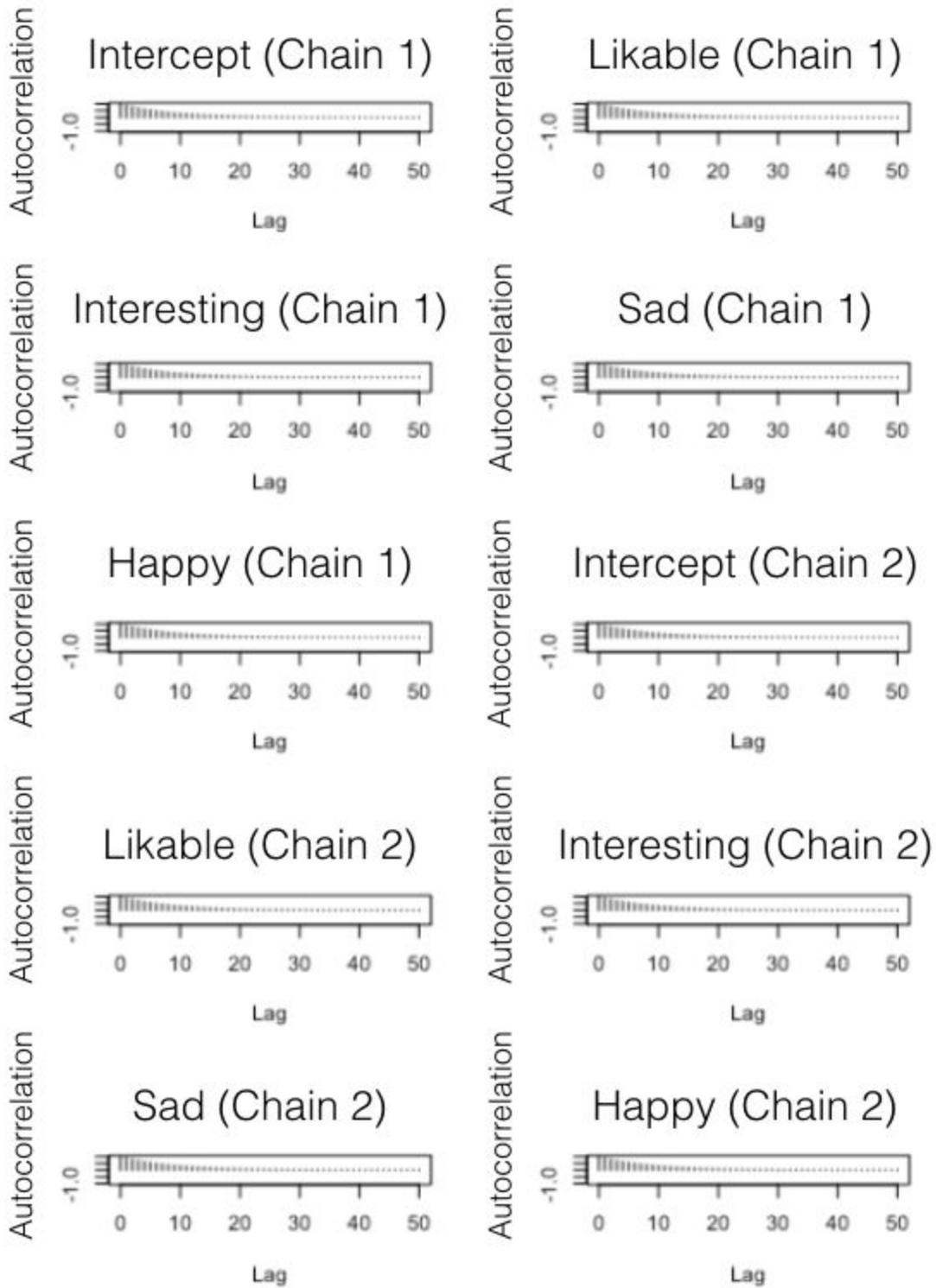

Fig. S6. Autocorrelation plot for Pre-diagnosis model MCMC simulations.

## VI. Frequentist Logistic Regression

```
                           Logit Regression Results
==============================================================================
Dep. Variable:                 target   No. Observations:                24713
Model:                          Logit   Df Residuals:                    24703
Method:                           MLE   Df Model:                            9
Date:                Mon, 11 Jul 2016   Pseudo R-squ.:                 0.01383
Time:                        17:40:18   Log-Likelihood:                -16832.
converged:                       True   LL-Null:                       -17068.
                                        LLR p-value:                 5.088e-96
==============================================================================
```

| All-data      | coef    | std err | z      | P>\|z\| | [95.0% Conf. Int.] |        |
|---------------|---------|---------|--------|---------|--------------------|--------|
| const         |  0.4996 |  0.078  |  6.374 |  0.000  |  0.346             |  0.653 |
| hue           |  0.3228 |  0.084  |  3.861 |  0.000  |  0.159             |  0.487 |
| saturation    | -0.6014 |  0.087  | -6.880 |  0.000  | -0.773             | -0.430 |
| brightness    | -0.5709 |  0.098  | -5.836 |  0.000  | -0.763             | -0.379 |
| comment_count |  0.0619 |  0.008  |  7.893 |  0.000  |  0.047             |  0.077 |
| like_count    | -0.0044 |  0.000  | -9.130 |  0.000  | -0.005             | -0.003 |
| url           |  0.0373 |  0.013  |  2.822 |  0.005  |  0.011             |  0.063 |
| has_filter    | -0.1232 |  0.013  | -9.428 |  0.000  | -0.149             | -0.098 |
| has_face      |  0.2424 |  0.024  |  9.994 |  0.000  |  0.195             |  0.290 |
| face_ct       | -0.1048 |  0.018  | -5.859 |  0.000  | -0.140             | -0.070 |

```
==============================================================================
Dep. Variable:                 target   No. Observations:                18513
Model:                          Logit   Df Residuals:                    18503
Method:                           MLE   Df Model:                            9
Date:                Wed, 13 Jul 2016   Pseudo R-squ.:                 0.01843
Time:                        22:54:47   Log-Likelihood:                -12065.
converged:                       True   LL-Null:                       -12291.
                                        LLR p-value:                 6.123e-92
==============================================================================
```

| Pre-diagnosis | coef    | std err | z       | P>\|z\| | [95.0% Conf. Int.] |        |
|---------------|---------|---------|---------|---------|--------------------|--------|
| const         | -0.4615 |  0.093  |  -4.949 |  0.000  | -0.644             | -0.279 |
| hue           |  0.7993 |  0.099  |   8.089 |  0.000  |  0.606             |  0.993 |
| saturation    | -0.2720 |  0.104  |  -2.621 |  0.009  | -0.475             | -0.069 |
| brightness    | -0.2105 |  0.117  |  -1.793 |  0.073  | -0.441             |  0.020 |
| comment_count |  0.0469 |  0.010  |   4.914 |  0.000  |  0.028             |  0.066 |
| like_count    | -0.0084 |  0.001  | -11.815 |  0.000  | -0.010             | -0.007 |
| url           | -0.0050 |  0.015  |  -0.324 |  0.746  | -0.035             |  0.025 |
| has_filter    | -0.0705 |  0.015  |  -4.802 |  0.000  | -0.099             | -0.042 |
| has_face      |  0.2674 |  0.029  |   9.158 |  0.000  |  0.210             |  0.325 |
| face_ct       | -0.2066 |  0.024  |  -8.563 |  0.000  | -0.254             | -0.159 |

Fig. S7. Frequentist logistic regression output. Frequentist and Bayesian regression output are largely in agreement.

## VII. Bayesian regression

### All-data model

|  | Depressed μ (σ) | Healthy μ (σ) | Coef μ (σ) | HPD Level | HPD Interval | |
|---|---|---|---|---|---|---|
| Intercept |  |  | .500 (.080) | 99% | .298 | .710 |
| Hue | .345 (.162) | .338 (.157) | .325 (.085) | 99% | .114 | .547 |
| Saturation | .338 (.157) | .347 (.155) | -.602 (.087) | 99% | -.828 | -.382 |
| Brightness | .535 (.138) | .547 (.145) | -.572 (.100) | 99% | -.835 | -.321 |
| Comments | 1.077 (2.150) | .992 (2.013) | .062 (.008) | 99% | .042 | .083 |
| Likes | 16.168 (34.874) | 18.939 (34.214) | -.004 (.005) | 99% | -.006 | -.003 |
| Posts/day | 1.875 (1.961) | 1.667 (1.775) | .037 (.013) | 99% | .005 | .072 |
| Has filter | .829 (1.108) | .871 (1.524) | -.124 (.013) | 99% | -.157 | -.089 |
| Has face | .769 (1.137) | .615 (.882) | .243 (.024) | 99% | .181 | .305 |
| Face count | .631 (.897) | .623 (.984) | -.105 (.018) | 99% | -.151 | -.059 |

Table S2. Logistic regression output for All-data model1 (N=24,713). HPD Level = Highest Posterior Density Level, the probability that a regression coefficient falls within the given HPD Interval. HPD Levels listed are highest probabilities with which it can be claimed that a coefficient's HPD Interval excludes zero.

### Pre-diagnosis model

|  | Depressed μ (σ) | Healthy μ (σ) | Coef μ (σ) | HPD Level | HPD Interval | |
|---|---|---|---|---|---|---|
| Intercept |  |  | .463 (.093) | 99% | -.695 | -.211 |
| Hue | .360 (.166) | .338 (.157) | .802 (.099) | 99% | .545 | 1.054 |
| Saturation | .348 (.157) | .347 (.155) | -.271 (.104) | 99% | -.522 | -.002 |
| Brightness | .534 (.136) | .547 (.145) | -.209 (.118) | 90% | -.410 | -.026 |
| Comments | .912 (1.771) | .992 (2.013) | .047 (.010) | 99% | .023 | .072 |
| Likes | 12.719 (28.912) | 18.939 (34.214) | -.008 (.001) | 99% | -.010 | -.007 |
| Posts/day | 1.877 (1.931) | 1.667 (1.775) | -.004 (.015) | 30% | -.012 | .000 |
| Has filter | .907 (1.191) | .871 (1.524) | -.071 (.015) | 99% | -.108 | -.030 |
| Has face | .743 (1.030) | .615 (.882) | .267 (.029) | 99% | .189 | .340 |
| Face count | .57 (.824) | .623 (.984) | -.207 (.024) | 99% | -.268 | -.144 |

Table S3. Logistic regression output for Pre-diagnosis model ( N=18,513). HPD Level = Highest Posterior Density Level, the probability that a regression coefficient falls within the given HPD Interval. HPD Levels listed are highest probabilities with which it can be claimed that a coefficient's HPD Interval excludes zero.

A posterior predictive check showed that All-data observations replicated from the joint posterior distribution consistently overestimated the proportion of depressed observations (replicated: 53.5% depressed; original: 30.9%), with a p-value of $1.0^8$. Pre-diagnosis observations sampled from the joint posterior distribution slightly underestimated the proportion of depressed observations (replicated: 30.02% depressed; original: 37.97%), with a posterior predictive p-value of 0.039. Gelman et al. (49) suggested that a model with good replication accuracy should generate posterior predictive p-values within the range of 0.05-0.95. Note that an extreme posterior predictive p-value does not mean that a model is wrong, just that it fails to be "right enough" to render a reasonable replication of its input. All models nevertheless far outperformed a simple null model in the capacity to correctly predict class membership.

| | Depressed μ (σ) | Healthy μ (σ) | Coef μ (σ) | HPD Level | HPD Interval | |
|---|---|---|---|---|---|---|
| Intercept | | | -2.374 (.175) | 20% | -.045 | -.004 |
| Happy | 2.300 (1.042) | 2.511 (1.109) | -.193 (.034) | 99% | -.279 | -.105 |
| Sad | .840 (.598) | .757 (.614) | .100 (.039) | 95% | .024 | .176 |
| Likable | 2.393 (.918) | 2.514 (.952) | .027 (.050) | 35% | .007 | .052 |
| Interesting | 2.316 (.816) | 2.367 (.859) | .041 (.041) | 65% | .003 | .080 |

Table S4. Logistic regression output for Ratings model (N=8,976). HPD Level = Highest Posterior Density Level, the probability that a regression coefficient falls within the given HPD Interval. HPD Levels listed are highest probabilities with which it can be claimed that a coefficient's HPD Interval excludes zero.

A posterior predictive check of Ratings model showed that sample observations replicated from the joint posterior distribution accurately represented the true proportion of depressed observations (replicated: 44.2% depressed; original: 43.9%), with a posterior predictive p-value of 0.516.

---

[8] In the context of logistic regression, the posterior predictive p-value assesses the frequency with which samples drawn from the simulated posterior overpredicts reference class membership, compared to reference class prevalence in the original data.

VIII. Instagram filter examples

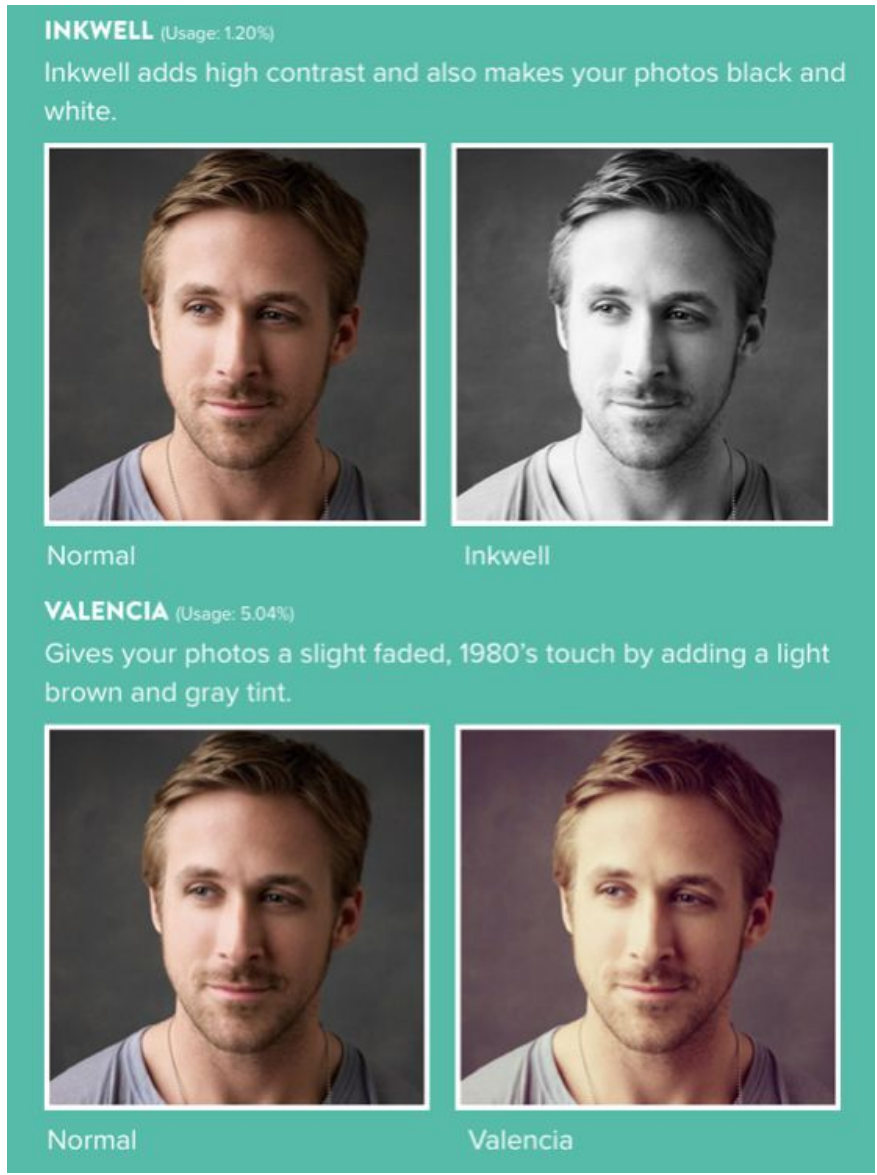

Fig. S8. Examples of Inkwell and Valencia Instagram filters. Inkwell converts color photos to black-and-white, Valencia lightens tint. Depressed participants most favored Inkwell compared to healthy participants, Healthy participants most favored Valencia compared to depressed participants.
Image credit: filterfakers.com

## IX. Correlation tables

### All-data model

|  | Hue | Satur. | Bright. | Comm. | Likes | Posts | Filter | Face |
|---|---|---|---|---|---|---|---|---|
| Saturation | .17 | | | | | | | |
| Brightness | -.22 | -.28 | | | | | | |
| Comments | -.07 | -.05 | .10 | | | | | |
| Likes | -.09 | -.10 | .17 | .55 | | | | |
| Posts | .05 | .03 | -.06 | -.02 | -.03 | | | |
| Has filter | .12 | .08 | -.02 | -.05 | -.07 | .56 | | |
| Has face | .03 | .04 | -.05 | -.00 | -.02 | .69 | .34 | |
| Face count | .01 | .04 | -.03 | .02 | .00 | -.02 | -.01 | .42 |

Table S5. Pearson's product-moment correlation scores for All-data model features.

### Pre-diagnosis model

|  | Hue | Satur. | Bright. | Comm. | Likes | Posts | Filter | Face |
|---|---|---|---|---|---|---|---|---|
| Saturation | .16 | | | | | | | |
| Brightness | -.22 | -.27 | | | | | | |
| Comments | -.05 | -.06 | .14 | | | | | |
| Likes | -.08 | -.13 | .23 | .49 | | | | |
| Posts | .06 | .03 | -.06 | -.04 | -.06 | | | |
| Has filter | .13 | .09 | -.03 | -.04 | -.08 | .63 | | |
| Has face | .04 | .05 | -.05 | -.01 | -.05 | .66 | .39 | |
| Face count | .01 | .03 | -.02 | .03 | -.01 | -.02 | -.02 | .45 |

Table S6. Pearson's product-moment correlation scores for Pre-diagnosis model features.

### Ratings model

|  | Happy | Sad | Likable | Interest. |
|---|---|---|---|---|
| Sad | -.41 | | | |
| Likable | .79 | -.29 | | |
| Interesting | .53 | -.09 | 0.77 | |
| Hue | .02 | -.02 | -.01 | -.03 |
| Saturation | .02 | -.07 | -.02 | -.04 |
| Brightness | .05 | -.04 | .04 | .03 |
| Posts | -.02 | .04 | -.01 | .02 |
| Comments | .00 | .02 | -.02 | -.03 |
| Likes | .04 | -.02 | .05 | .06 |
| Has filter | .03 | .00 | .02 | .01 |
| Has face | .16 | .05 | .06 | .00 |
| Face count | .25 | -.10 | .11 | .02 |

Table S7. Pearson's product-moment correlation scores for Ratings model features (columns) with ratings and computational features (rows).

X. Ratings inter-rater agreement

Rater agreement was measured by randomly selecting two raters from each photo, and computing Pearson's product-moment correlation coefficient from the resulting vectors. To mitigate sampling bias, we ran a five-fold iteration of this process and averaged the resulting coefficients. Rater agreement showed positive correlations across all ratings categories ($p < 1e-38$ for all values shown): $r_{happy} = .39$, $r_{sad} = .19$, $r_{interesting} = .17$, $r_{likable} = .27$

XI. Differences in filter use, Pre-diagnosis model

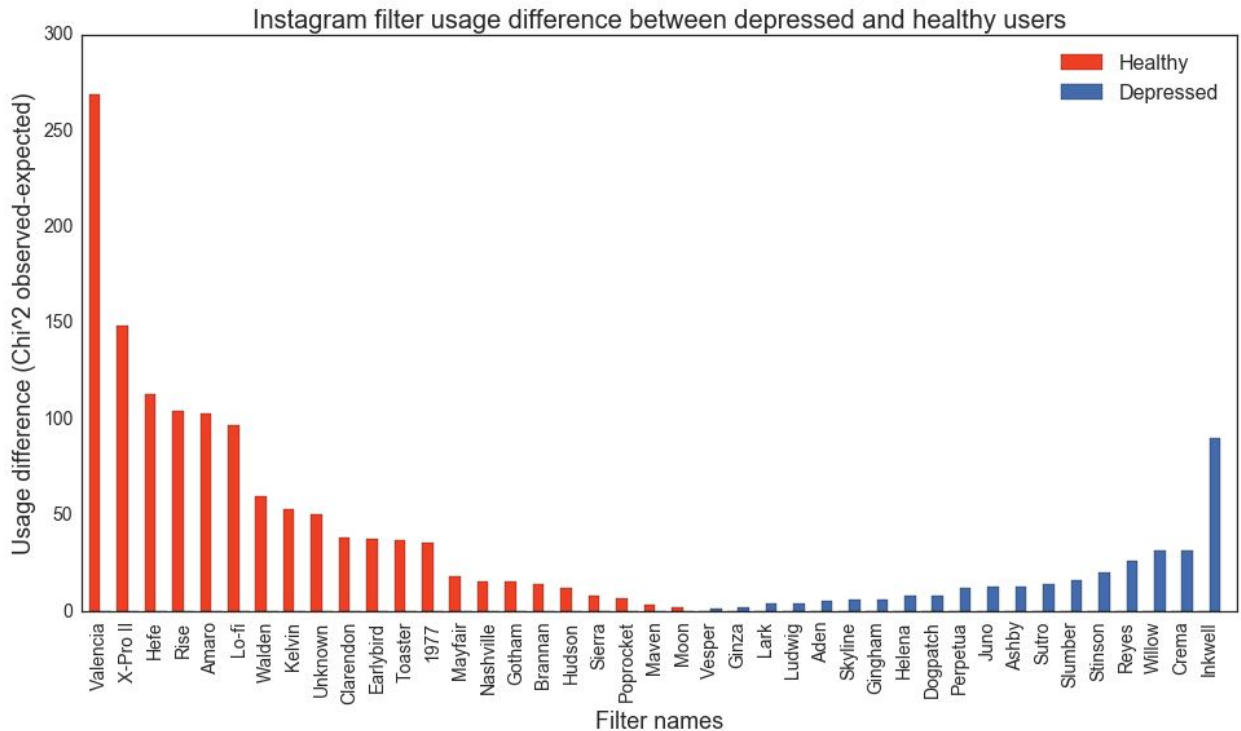

Fig. S9. Instagram filter usage among depressed and healthy participants. Bars indicate difference between observed and expected usage frequencies, based on a Chi-squared analysis of independence. Blue bars indicate disproportionate use of a filter by depressed compared to healthy participants, orange bars indicate the reverse. Pre-diagnosis model results are displayed, similar results were observed for the All-data model, see main text Figure 3.